\documentclass[twocolumn]{aastex631}

\shortauthors{Vipin Sudevan \& Pisin Chen}

\usepackage{apjfonts}
\usepackage[symbols,nogroupskip,nonumberlist]{glossaries-extra}
\shorttitle{${\tt PUREPath}$}

\begin{document}

\title{PUREP\MakeLowercase a\MakeLowercase t\MakeLowercase h: A Deep Latent Variational Model for Estimating CMB Posterior over Large Angular Scales of the Sky}
\author{Vipin Sudevan}
\affiliation{Leung Center for Cosmology and Particle Astrophysics, National Taiwan University, Taipei 10617, Taiwan; vipinsudevan1988@gmail.com}

\author{Pisin Chen}
\affiliation{Leung Center for Cosmology and Particle Astrophysics, National Taiwan University, Taipei 10617, Taiwan}
\affiliation{Department of Physics and Center for Theoretical Sciences, National Taiwan University, Taipei 10617, Taiwan}
\affiliation{Graduate Institute of Astrophysics, National Taiwan University, Taipei 10617, Taiwan}
\affiliation{Kavli Institute for Particle Astrophysics and Cosmology, SLAC National Accelerator Laboratory, Stanford University, Stanford, California 94305, USA}

\begin{abstract}
{We present a comprehensive neural architecture, the ${\tt PUREPath}$, 
which leverages a nested  
{\bf P}robabilistic multi-modal {\bf U}-Net framework, augmented by the inclusion of probabilistic {\bf R}esNet 
blocks in the {\bf E}xpanding {\bf Path}way of the decoders, to estimate the posterior density of the 
Cosmic Microwave Background (CMB) signal conditioned on the observed CMB data 
and the training dataset. 
By seamlessly integrating   
Bayesian statistics and variational 
methods our model effectively  minimizes foreground contamination 
in the observed CMB maps.  
The model is trained using  
foreground and noise contaminated CMB temperature maps simulated at  
Planck LFI and HFI frequency channels 30 - 353 GHz using publicly available 
Code for Anisotropies in 
the Microwave Background (CAMB) and Python Sky Model (PySM) packages. 
During training, our model transforms initial 
prior distribution on the model parameters  
to posterior distributions based on the training data.
From the joint full posterior of the model parameters, 
during inference, a predicitve CMB posterior and summary statistics such as the
predictive mean, variance etc of the cleaned CMB map is estimated. 
The predictive standard deviation map provides a direct and 
interpretable measure of uncertainty per pixel in the predicted mean CMB map. 
The cleaned CMB map 
along with the error estimates can be used for more accurate measurements 
of cosmological parameters and other cosmological
analyses.
}
\end{abstract}


\section{Introduction}
\label{into}
The CMB fluctuations provide us valuable insights regarding the origin, geometry,  
and composition  of our  
universe while rendering  
stringent constraints on cosmological parameters~\cite{Planck:2018vyg}. 
Several ground~\citep{Hincks_2010,Li:2017drr,Hui:2018cvg,CMB-S4:2022ght}, 
ballon~\citep{Masi:2002hp,2014SPIE.9153E..1LL,SPIDER:2017xxz} and 
satellite~\citep{2003ApJS..148....1B, Planck:2013oqw,LiteBIRD:2022cnt,Adak:2021lbu} 
-based scientific missions have already observed or 
are at various stages of either planning or observing these fluctuations with 
increasingly higher sensitivities. 
However, strong microwave emissions from various galactic 
and extra-galactic astrophysical sources, 
the foregrounds,  and inherent instrumental noise of the detectors 
contaminates the observed CMB signal. It is crucial for the success of any 
CMB mission to either disentangle or mitigate these 
foregrounds and detector-noise contributions to   
disinter the physics encoded in the CMB fluctuations.

Recent decades saw development of several foreground minimization and component 
separation methods which can be broadly classified into $parametric$ and $non-parametric$ approaches. 
$Parametric$ 
methods like Commander (\cite{Eriksen2008}, \cite{Eriksen2008a}), 
Template-fitting (\cite{Land:2005cg}, \cite{Jaffe:2006fh}) requires intricate modeling of 
either CMB and/or foregrounds to separate the CMB signal in the observed maps. On 
the contrary, $non-parametric$ methods makes  
minimal assumptions concerning the nature of the foregrounds,  and CMB 
while estimating the CMB signal. Some of the prominent $non-parametric$ methods includes 
internal-linear-combination (ILC)
(\cite{2003ApJS..148...97B}, \cite{2004ApJ...612..633E}, \cite{Tegmark2003}, \cite{2009A&A...493..835D}, \cite{Saha2017}) as well as 
Independent Component Analysis 
(ICA) method (\cite{Taylor:2006otn}, \cite{2013A&A...558A.118H}), among others. 

In recent years, due to remarkable achievements in the field of machine learning (ML)~\citep{5392560}  
in tackling 
various complex real-world challenges, ML is actively pursued   
to solve intricate problems in the realm of Physics. 
The image processing capabilities of ML models especually with the use of
convolutional layers,   
have spiked interest in the CMB community to 
invoke ML techniques for CMB analyses.  
Convolutional neural networks 
are developed to extract the CMB signal by minimizing the foregrounds in the  
CMB data~\citep{Petroff:2020fbf,Wang:2022ybb,Casas:2022teu,Yan:2023bjq, Yan:2024czw}, 
to inpaint masked regions  
in CMB maps~\citep{Yi:2020xgq,Montefalcone_2021},  
delensing in CMB polarization maps~\citep{Yan:2023oan}. ML models are 
also used to recognize different foreground models~\citep{Farsian:2020adf}, 
to remove E-B leakage in partial-sky~\citep{Pal:2022woh},   
estimation of B-mode signal at large angular scales after 
removing the foregrounds~\citep{Pal:2024cir}. 
\cite{Adams:2023uod} and~\cite{Petroff:2020fbf} developed   DeepShere~\citep{defferrard2020deepsphere} based U-Net~\citep{2015arXiv150504597R}
with concrete dropout~\citep{NIPS2017_84ddfb34} to minimize foreground contaminations in CMB data. 
Concrete dropout can be used to estimate a network's epistemic uncertainty 
due to insufficient training.   
These are some of the many instances where 
ML models are leveraged within CMB analyses.

In conventional ML framework, a model is trained to understand the relationship  
between the  inputs and  outputs by fixing the model parameters 
(${\omega}$), the weights and biases. 
In a deterministic neural network, these parameters remain unchanged 
when the network is used for predictions. 
They do not inherently take into account for either the epistemic 
 or aleotoric (due to intrinsic errors in the data) 
uncertainty and often leads to  situations where a trained  
network makes overly confident predictions on new unseen data.

Probabilistic neural networks~\citep{SPECHT1990109} like 
Bayesian neural  networks are developed to address this crucial shortcoming 
of the standard deep neural networks by incorporating uncertainty considerations 
in both the model and the data.   
Bayesian networks utilizes probabilistic layers in order to 
capture the uncertainty over weights or 
activations or both by incorporating prior knowledge 
of the parameters and propagating it through the network, 
thereby modeling the uncertainty in the predictions. 
This is achieved through Bayesian inference~\citep{Tipping2004}, 
where the posterior distribution over model parameters is 
updated given the training data 
by using Bayes' theorem. Since Bayesian posteriors are 
usually intractable,  
variational inference (VI)
techniques like variational Bayes (\cite{2013arXiv1312.6114K}, \cite{JMLR:v14:hoffman13a}) 
and stochastic gradient 
variational Bayes (\cite{2015arXiv150901631K}) are developed. 
These methods approximate 
the posterior distribution of the model parameters by optimizing 
a tractable surrogate distribution. 
Probabilistic layers in Bayesian networks 
provide a more intuitive way to understand the uncertainty associated 
with the model and its 
predictions. However, constructing and training a fully 
Bayesian neural network can be challenging and is nontrivial  
even in relatively simple applications. 

Our ${\tt PUREPath}$ is designed 
to minimize foreground contaminations in the observed full-sky CMB maps. The full-sky 
maps are projected into a plane and divided into  
$n_{\tt reg}$ distinct and disjoint regions as shown in 
Figure~\ref{inp_map} during pre-processing.
${\tt PUREPath}$ architecture is best characterised as a probabilistic, nested framework composed 
of $n_{\tt reg}$  probabilistic U-Net. Each U-Net 
consists of $n_{\tt map}$ encoders to facilitate a multi-modal learning,  
a latent space, and a decoder with probabilistic ResNets~\citep{2015arXiv151203385H}
 in the expanding 
pathway of each decoders at skip connections. 
The output of our model is  a distribution defined by MultivariateNormalDiag layer with inputs, 
the mean and scale$\_$diag,  as the decoder outputs.  
More details regarding our model is provided in Section~\ref{Architecture}.

\section{Formalism}
\label{formalism}

We briefly review how a Bayesian neural network which utilizes probabilistic 
layers is trained using Variational Inference.   

\subsection{Bayesian Neural Network}  
\label{BNN}

The parameters (${\omega}$) in a Bayesian network   are   
stochastic in nature and are sampled from respective posterior distributions 
$P(\omega | \mathcal{D})$. Here,  $\mathcal{D}$ is the 
training dataset with  inputs $\Bigl\{{\bf X}_j = \{{\bf x}_{i,\,j}\}_{i=1}^{n}\Bigl\}_{j=1}^{N}$, comprising of $N$
simulated sets of foreground contaminated CMB maps ${\bf x}_i$ at $n$ frequencies, 
and $N$ output ${\bf y}_j$ (the input simulated CMB map  
for the set $j$) i.e., 
\begin{equation}
\mathcal{D} = \Bigl\{\{{\bf x}_{i,\,j}\}_{i=1}^{n}, {\bf y}_j \Bigl\}_{j=1}^{N}
\end{equation} 
In the Bayesian approach, a prior distribution $P({\omega})$ of 
the model parameters $\omega$ is assumed 
based on the assumption as to which functions best generates the data. 
For a given $P({\omega})$ and $\mathcal{D}$, 
the posterior distribution $P(\omega | \mathcal{D})$ (\cite{graves2011practical}) 
is updated using the Bayes' principle as follows:  
\begin{equation}
P(\omega|\mathcal{D}) = \frac{P(\mathcal{D}|\omega)P({\omega})}{P(\mathcal{D})} \, .
\label{bayes}
\end{equation}
$P(\mathcal{D}|\omega) = \Pi_{i=1}^{N} P({\bf y}_i | {\bf X}_i, \omega)$ is  
$likelihood$ of the observing ${\bf y}_i$ 
given ${\bf X}_i$ and set of model parameters $\omega$. 

Accurate estimation of exact $P(\omega| \mathcal{D})$ using Bayesian inference 
is not practical as it requires integrating over 
all feasible values of $\omega$. 
A simpler alternative is    
to approximate the exact distribution with a more tractable 
surrogate or variational distribution $Q_\theta({\omega})$ with 
parameters $\theta$, the VI technique. 
These variational parameters ($\theta$) are optimized to 
minimize the difference between approximate and true posteriors. 
In practice, we minimize the the Kullback-Leibler (KL) 
divergence $\mathbb{D}_{KL} (Q_\theta({\omega}) || P({\omega} | \mathcal{D}))$~\citep{kullback1951information}, 
\begin{equation}
\mathbb{D}_{KL} (Q_\theta({\omega}) || P({\omega} | \mathcal{D})) = \mathbb{E}_Q [\log Q_\theta({\omega})] - \mathbb{E}_Q [\log P({\omega} | \mathcal{D})] + \log(P(\mathcal{D})) \, .
\label{kl}
\end{equation}
to find these variational parameters. The goal of optimization procedure is to 
determine the $\theta$ which minimizes KL-divergence, thereby 
facilitating a more practical approximation of the true posterior. 
During implementation, the usual approach is to maximize the so-called 
evidence lower bound (ELBO)~\citep{doi:10.1080/01621459.2017.1285773}, 
\begin{flalign}
ELBO(Q) & = \mathbb{E}_Q [\log P({\mathcal D} | \omega)] - \beta D_{KL} (Q_\theta(\omega) || P(\omega | {\mathcal D})) \, .
\label{ELBO}
\end{flalign}
Here, $\mathbb{E}_Q [\log P({\mathcal D} | \omega)]$ is the expected log-likelihood under 
the variational distribution $Q_\theta({\omega})$. $\beta$ is a tunable hper parameter 
to weight the KL-Diveregence. ELBO is preferred to train  Bayesian networks as it serves 
as a lower bound on the log marginal likelihood of the 
data and also measures 
the quality of the approximate posterior. 
The loss function ($\mathbb{L}$) is defined as the negative of ELBO,
\begin{equation}
\mathbb{L}_{\tt ELBO} = -ELBO(Q)\, .
\label{loss}
\end{equation}
During the training phase, this loss is 
minimized to find the optimal values of $\theta$ that defines the parameters of the distribution over weights. 
The distribution $Q_\theta(\omega)$ can be described in terms 
of weight perturbations $\omega = \bar{\omega} + \Delta \omega$, where $\bar{\omega}$ and $\Delta \omega$ are 
the mean weights 
and a stochastic perturbation for ${\omega}$ respectively, $\theta = \{\bar{\omega}, 
\Delta\omega\}$. 

\begin{figure}
 \centering
\includegraphics[scale=0.18]{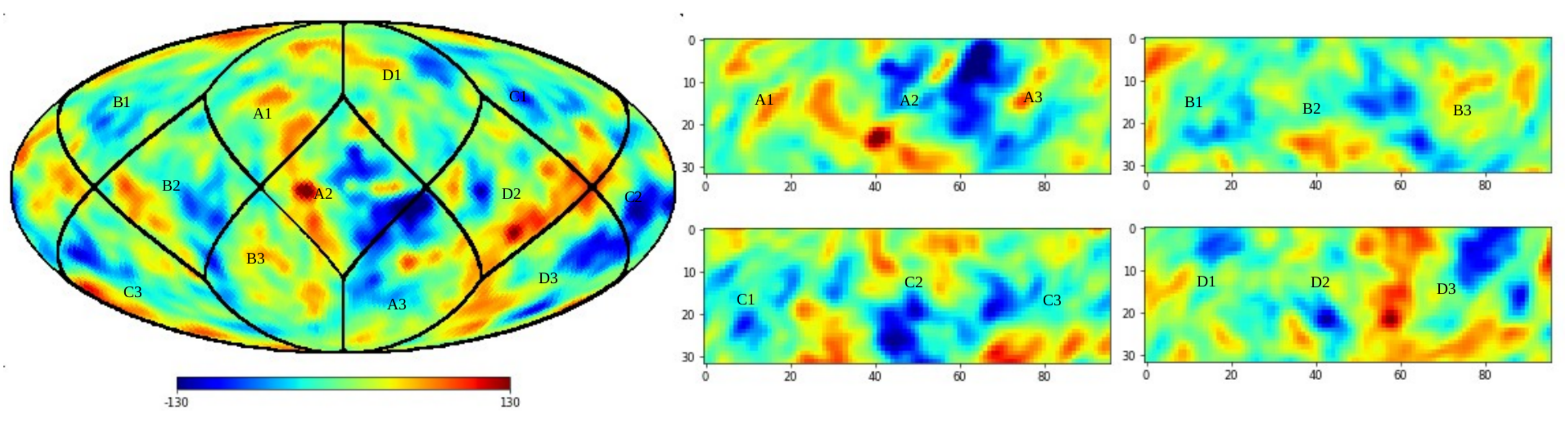}
\caption{In the left panel we show a simulated CMB map using HealPix Mollewide 
projection. The map is divided with solid black lines into 12 equal area regions. 
These 12 regions are  of dimension $N_{\tt side}\, \times\, N_{\tt side}$ 
and are stitched 
together to form the four plane images of dimension $N_{\tt side}\, \times\, 3N_{\tt side}$.   The choice of the regions in the planar maps is such that 
the regions which are in contact in the planar maps are 
also in contact in the real sky and the dimension of all 
images is same.}   
\label{inp_map}
\end{figure}

\subsection{Weight Perturbations with Flipout}

We follow a weight perturbation technique known as 
the Flipout~\citep{2018arXiv180304386W} to train our network. This method is based on     
two key assumptions regarding the nature of perturbations $\Delta \omega$, the 
perturbations of different weights are independent to each other 
and their distribution is symmetric around zero.  
Under these assumptions, the distribution of 
the perturbations is invariant to an element-wise 
multiplication by a random sign matrix. Flipout introduces  a common base perturbation 
$\Delta\hat{\omega}$ during 
training, which is shared 
by all training examples in a mini-batch. For each individual example, 
the corresponding weight perturbations are achieved as follows: 
\begin{equation}
\Delta \omega = \Delta\hat{\omega} \circ r_ns_n^T \, .
\end{equation}
Here, the  subscript $n$ denotes the index of the data-points in a mini-batch. 
$r_n$ and $s_n$ are entries 
of random vectors uniformly sampled from $\pm 1$. Flipout technique 
introduces different weight perturbations 
for each example in a mini-batch thus ensuring the gradients to be  
decorrelated between different 
training examples in a mini-batch. This significantly reduce the variance 
in the gradient updates 
when averaging over a mini-batch.

For Bayesian networks, $\bar{\omega}$ and $\Delta\hat{\omega}$ are 
obtained via standard backpropogation 
with stochastic 
optimization algorithms. For mini-batch optimization, the batch loss is written as
\begin{equation}
\mathbb{L}_i = \beta D_{KL} (Q_\theta({\omega}) || P(\theta)) - \frac{1}{N} \sum_{n=1}^{N} \log P({\mathcal D}_i^n|{\bf \omega}^{i})    \, ,
\end{equation}
where $i$ iterates over the mini-batch indices, $n$ and $N$ represents the index 
corresponding to the particular example in the mini-batch and total size of 
each mini-batch respectively.  
Even though only one set of parameters ${\omega}^{i}$
is drawn from $Q_\theta({\omega})$ for each mini-batch, Flipout  
ensures the parameters are different for each individual examples ${\mathcal D}_i^n$. 
\subsection{Estimating CMB Posterior}
After training, the model provides a variational posterior $Q_\theta({\omega|\theta^\ast})$, 
where $\theta^\ast$ are the optimized values of the parameters of the variational 
distribution.  
The predictive CMB posterior ${\hat P}({\bf y}^\ast|{\bf X}^\ast, \mathcal{D})$ 
of the estimated cleaned CMB map (${\bf y}^\ast$) 
conditioned on the input maps 
(${\bf X}^\ast$) and the training dataset ($\mathcal{D}$), 
is estimated using the variational posterior 
$Q_\theta({\omega|\theta^\ast})$ 
of the model parameters as follows, 
\begin{align}
{\hat P}({\bf y}^\ast|{\bf X}^\ast, \mathcal{D}) &= \int {\hat P}({\bf y}^\ast|{\bf X}^\ast, \omega) P(\omega|\mathcal{D}) d\omega  \nonumber \\ &\approx \int {\hat P}({\bf y}^\ast|{\bf X}^\ast, \omega) \mathcal{Q}_\theta(\omega|\theta^\ast) d\omega \, .
\label{samp}
\end{align}
This integral can be numerically evaluated using Monte Carlo (MC) sampling as:
\begin{equation}
{\hat P}({\bf y}^\ast |{\bf X}^\ast, \mathcal{D}) \approx \frac{1}{S} \sum_{s=1}^{S} {\hat P}({\bf y}^\ast | {\bf X}^\ast, \omega^s)  \, , 
\label{posterior}
\end{equation}
where, ${\hat P}({\bf y}^\ast | {\bf X}^\ast, \omega^s)$ is a predictive distribution 
corresponding to the model parameters $\omega^s$ sampled from  $Q_\theta({\omega|\theta^\ast})$  i.e., $\omega^s\,\sim\,Q_\theta({\omega|\theta^\ast}$).   
$s$ and $S$ in Eqn.~\ref{posterior}	represents the index of the sampling 
step and the total number of MC samplings respectively. 
${\hat P}({\bf y}^\ast | {\bf x}^\ast, \omega^s)  = \mathcal{N}({\bf y}^\ast | \mu_{\tt CMB}({\bf x}^\ast, \omega^s), \sigma_{\tt CMB}({\bf x}^\ast, \omega^s))$~\citep{abdar2021review}
where  $\mu_{\tt CMB}$ and $\Sigma_{\tt CMB}$ are 
the predictive mean and variance estimated from the samples generated in  
the step $s$.

\subsection{Estimation of Uncertainty}
\label{uncertainty}
The advantage of implementing a Bayesian network is its ability  to quantify 
both 
epistemic and aleatoric uncertainty. From the predictive CMB posterior 
distribution ${\hat P}({\bf y}^\ast |{\bf x}^\ast, \mathcal{D})$, 
we can compute summary statistics such as the predictive mean, variance 
and other higher-order statistics of the cleaned CMB map.
To quantify the uncertainty in the predictions, we estimate the predictive mean,  
variance and standard deviation etc as follows:
\begin{align}
\mathbb{E}[{\bf y}^\ast | {\bf x}^\ast, {\mathcal D}] &\approx \frac{1}{S} \sum_{s=1}^{S} \mu_{\tt CMB}^s \label{pmean}\, ,\\
Var[{\bf y}^\ast |{\bf x}^\ast, \mathcal{D}]  &\approx \frac{1}{S} \sum_{s=1}^{S}\Bigl((\mu_{CMB}^s - \mathbb{E}[{\bf y}^*])^2 + (\sigma_{\tt CMB}^s)^2 \Bigl)\label{pvar}\, , \\
\sigma[{\bf y}^\ast |{\bf x}^\ast, \mathcal{D}] &= \sqrt{Var[{\bf y}^\ast |{\bf x}^\ast, \mathcal{D}]} \label{pstd}\, .
\end{align}


\section{Network Architecture}
\label{Architecture}

\begin{figure}
\includegraphics[scale=0.35]{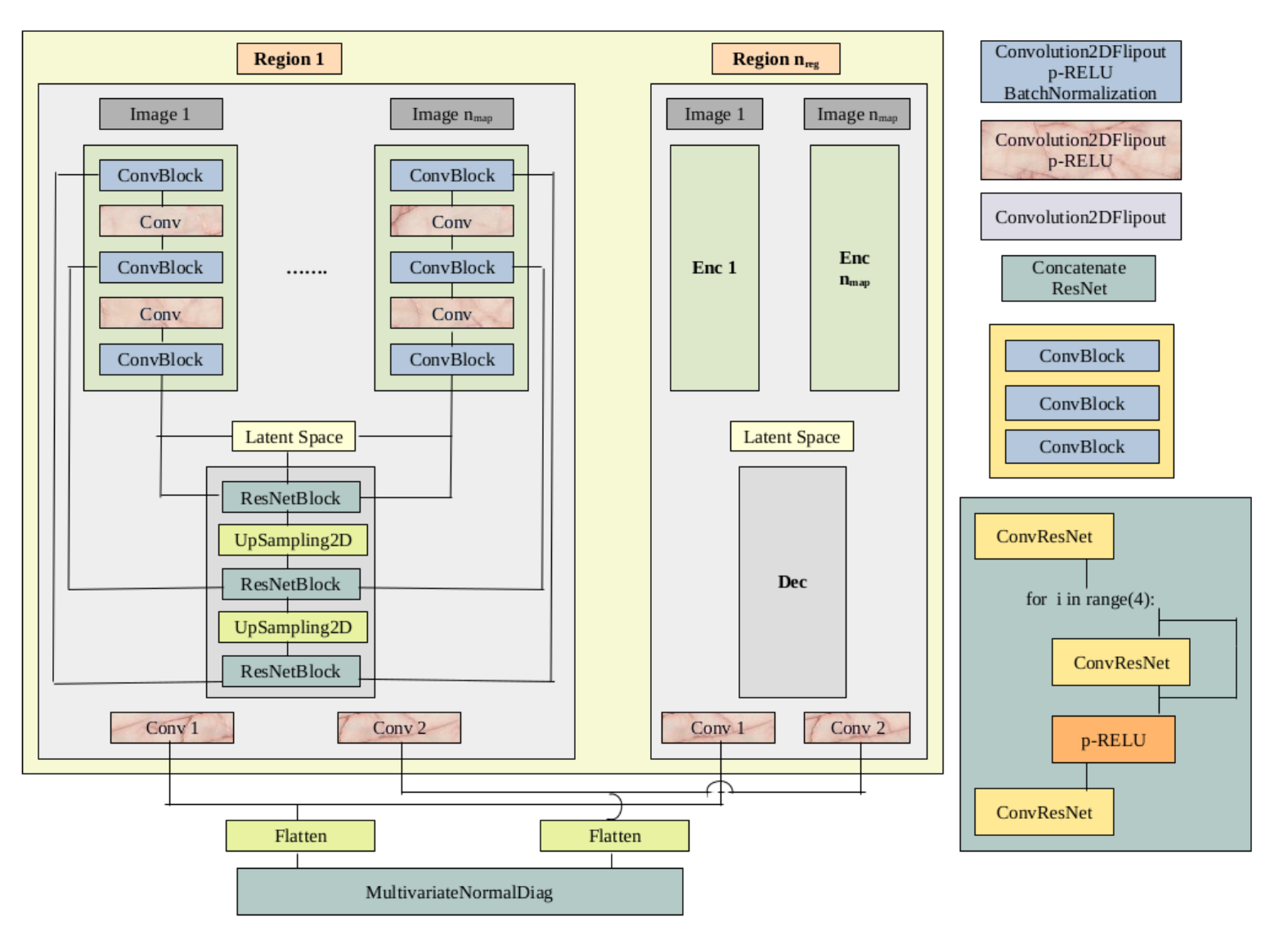}
\label{net}
\caption{A pictorial representation of our ${\tt PUREPath}$ Model. Each Region has its own multi-encoder-latent-decoder
units. Each encoder block in turn has separate encoder units to process each input map 
 in that region merging into a single 
latent space  and finally to a single Decoder unit. All the output feature maps 
finally goes to a MultivariateNormalDiag layer, the output of our model. }
\label{net_pic}
\end{figure}

The ${\tt PUREPath}$'s hybrid architecture is shown in  Figure~\ref{net_pic}.   
Depending on the total number of regions $N_{\tt reg}$ our network has that many U-Nets to 
process images corresponding to each regions. 
All the feature extraction layers in the U-Nets and ResNets are 
Tensorflow Probability~\citep{tensorflow2015-whitepaper}  
Convlution2DFlipout layers with kernel size $3\times5$ and 
strides 1 and with {\it parametric}-RELU ({\it p}-RELU) activation. 
They are initialized using default multivariate normal 
distributions with random mean and 
standard deviation and mean-field normal functions provided by 
Tensorflow as prior and posterior respectively. 
We perform a strided convolution to downsample the image with stride set to 2 in 
the Convoultion2DFlipout followed by {\it p}-ReLU activation.  The number 
of filters for all the convolution layers in the encoder is fixed at 16. 
The feature maps from the final layer of all the encoders in a U-net are combined 
together by performing an element-wise addition along the channels axis 
to from a high-dimensional, abstract representation of the input images in the 
latent space. These represenations are passed to decoder along with the 
encoder outputs. The decoder 
consists of ResNet Block in its expanding pathway at every skip-connection. In our formulation 
of ResNet (refer to Figure~\ref{net_pic}) three back-to-back convolutions 
operations are performed each 
followed by an activation with p-RELU and BatchNormalization. After these 
operations, the input is added back to 
the new feature maps to form the residual connection. 
The three consecutive convolution operations within each ResNet block  uses Convolution2DFlipout 
with kernel sizes $1\times 1$, $3\times 5$ and $1\times 1$. 
This process is repeated total 6 times. The three ResNet blocks in the decoder has 
input and ouput convolution filters set at (16, 16), (16, 32) and final block 
has (32, 32).  The feature maps from the final ResNet block is sent to two 
Convolution2DFlipout layers with p-ReLU activation and kernel size (1,1), thereby 
allowing each decoder to generate two feature maps at original input 
image dimension at the 
output.  
The final layer of our network is a distribution defined 
by a `MultivariateNormalDiag' layer parametrized by two inputs: `${\tt loc}$' and 
`${\tt scale\_diag}$'. 
The inputs are obtained by concatenating corresponding feature map from 
all the decoders and flattening them to get arrays of length $N_{\tt pix}$, where 
$N_{\tt pix}$ is the total number of pixels in the full-sky map as shown in Figure~\ref{inp_map}. 
This layer introduces a probabilistic 
element to the output and the mean of the distribution 
is the predicted output while 
the standard deviation is interpreted 
as the model's uncertainty in the prediction, 
thereby offering insights into the 
robustness and reliability of the model's output.
 
\section{Data Simulations}
\label{simulations}
We train our network using foreground 
and noise contaminated CMB 
maps at several Planck frequency channels simulated using 
publicly available software packages 
CAMB~\citep{2011ascl.soft02026L}, 
HealPix~\citep{Gorski:2004by}, and PySM~\cite{Thorne:2016ifb}. 
CAMB and PySM packages are employed to generate 1200 full-sky CMB simulations  
 based on the $\Lambda$ cold 
dark matter framework. We sample the following 
cosmological parameters $H_{0}$, $\Omega_b$, $\Omega_c$, $\tau$, $A_s$, $n_s$ 
 from their respective  
Gaussian distribution $\mathcal{N}(m_P, \sigma_P^2)$, 
with $m_P$ set to the best-fit value provided by~\cite{Planck:2018vyg} and 
with a 1$\sigma$ standard deviation. 
Employing CAMB, we 
generate a  lensed CMB power spectrum, corresponding to the sampled set of 
cosmological parameters, and the 
full-sky map with  PySM.

Foreground emission maps are simulated using PySM at 7 Planck 
frequencies i.e., Planck LFI 30, 44, 70 GHz, and Planck HFI 100, 
143, 217 and 353 GHz. The dominant sources of contaminators we consider 
are: 
thermal dust, synchrotron, free-free and anomolous microwave emission (AME). 
For thermal dust we use `d1 model' provided by PySM, while 
`s1 model' is employed to simulate synchrotron emissions. Free-free and 
AME simulations are generated using `f1' and `a2' model respectively. 
To train our network efficiently, we  
simulate 1200 realizations of these foregrounds at all frequencies of interest    
by sampling amplitude $A_{\tt foreg}$ and spectral index $\beta_{\tt foreg}$  of 
these foregrounds from  
a Gaussian distribution. The mean of this distribution is the corresponding 
$A_{\tt foreg}$ or $\beta_{\tt foreg}$ template provided by PySM and 
a standard deviation of 10$\%$ from the mean value.  

To create 1200 noise realizations for the Planck LFI and HFI maps, we 
use the intensity noise variance values given in the fifth column of
the respective frequency band Planck map files. Noise is assumed to be Gaussian, 
isotropic 
and uncorrelated between pixel to pixel. 
All the CMB, foregrounds and
Noise maps are generated  at HealPix $N_{\tt side} = 32$ 
with a Gaussian smoothing of FWHM $1.83^{\circ}$ ($\approx$ size of pixel in a 
HealPix map of $N_{\tt side} = 32$). 
We co-add 1200 samples of CMB, four different foregrounds and noise 
realizations corresponding to each of the 7 Planck frequency channels, 
to generate 1200 sets of simulated foreground contaminated CMB maps at each frequency.

\subsection{Data Preprocessing}
\label{preprocessing}
The spherical full-sky maps are transformed into approximate 
plane images~\citep{Wang:2022ybb} by 
reoredering all the maps 
from native HealPix Ring format to Nested pixellation 
scheme and then divide the resulting Nested maps 
into 12 equal area regions as shown in Figure~\ref{inp_map}. 
Each of these pieces 
are of dimension ($N_{\tt side} \times N_{\tt side}$). 
After this, we 
arrange three pieces together forming 4 independent 
($N_{\tt side} \times 3N_{\tt side}$) planar maps. 
This procedure results in 4 sets of 1200 planar images corresponding to 
each Planck frequency channel 
and the input CMB.  
We take 1100 samples as the training and remaining 100 as testing 
dataset. 15$\%$ of the training dataset is set as validation dataset. 

\section{Methodology}
\label{methodology}
We initialize the network with using the default prior and posterior distributions provided 
by Tensorflow. The  training data is 
provided to the network in batches of 16 samples per batch. 
This vastly reduces  
memory consumption and smaller batches can act 
as a form of regularization by introducing  a level 
of uncertainty to the weight updates.  
We have set the tunable hyper parameter $\beta$ to 1/10 in our analysis to keep both 
negative log likelihood and KL-Divergence at similar scales.
The network is trained using VI as discussed in the 
Section~\ref{formalism}. We use Adam optimization scheme~\citep{2014arXiv1412.6980K} initialized 
with learning rate set at 0.001. The learning rate is gradually reduced by 25$\%$ 
if the validation loss is not improved over consecutive 50 training epochs. 
We set a lower bound 
for the learning rate as $10^{-6}$. Our network stops training if the validation 
loss is not improved over a consecutive 200 epochs or if learning rate 
reaches its lower bound. 
We save the set of mean and standard deviation parameters of 
the model weight distributions corresponding to the lowest validation loss.

During inference,  to estimate the CMB posterior, we sample model parameters  
$\{\omega_i\}_{i=1}^{N}$, where $N$ is the total number of samples, from  
the learned variational distribution $Q_\theta({\omega|\theta^\ast})$. 
For each sampled set of $\omega_i$  we generate a predictive distribution for the CMB map 
${\hat P}({\bf y}|{\bf X}, \omega_i)$ given ${\bf X}$ and $\omega_i$. 
${\hat P}({\bf y}|{\bf X}, \omega_i)$ represents our model's uncertainty about the cleaned 
CMB map, accounting for the aleotoric uncertainty.  We approximate 
a predictive CMB posterior distribution  ${\hat P}({\bf y}|{\bf X}, \mathcal{D})$ as: 
\begin{equation}
{\hat P}({\bf y}|{\bf X}, \mathcal{D}) \approx \frac{1}{N} \sum_{i=1}^{N} 
{\hat P}({\bf y}|{\bf X}, \omega_i)
\end{equation}
Using the predictive mean and standard deviation maps corresponding to each realization 
of the network parameters $\omega_i$) we estimate the final predictive mean map, 
standard deviation map etc according the Eqns.~\ref{pmean} and~\ref{pstd}. 
The estimated standard deviation map effectively 
captures both the aleotoric and epistemic uncertainty in the cleaned CMB maps.

\section{Results from Observed Planck Maps}
\label{results}

\begin{figure*}
 \centering
\includegraphics[scale=0.185]{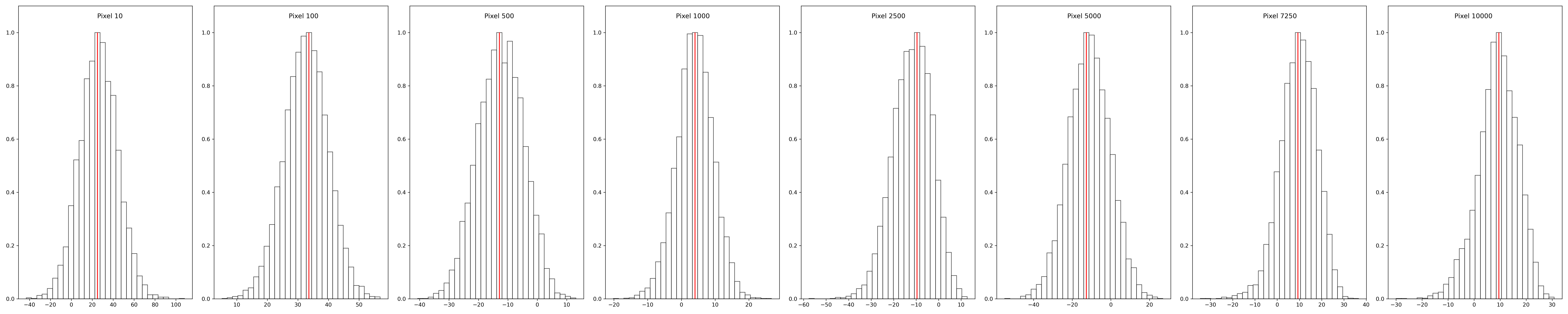}
\caption{We show the normalized densities of the CMB pixel temperatures for some 
selected pixels in black. The normalization for each density is
such that the peak corresponds to a value of unity. 
The position of mode value is shown by the red vertical lines.}
\label{post_data_pix}
\end{figure*}

\begin{figure}
\centering
\includegraphics[scale=0.225]{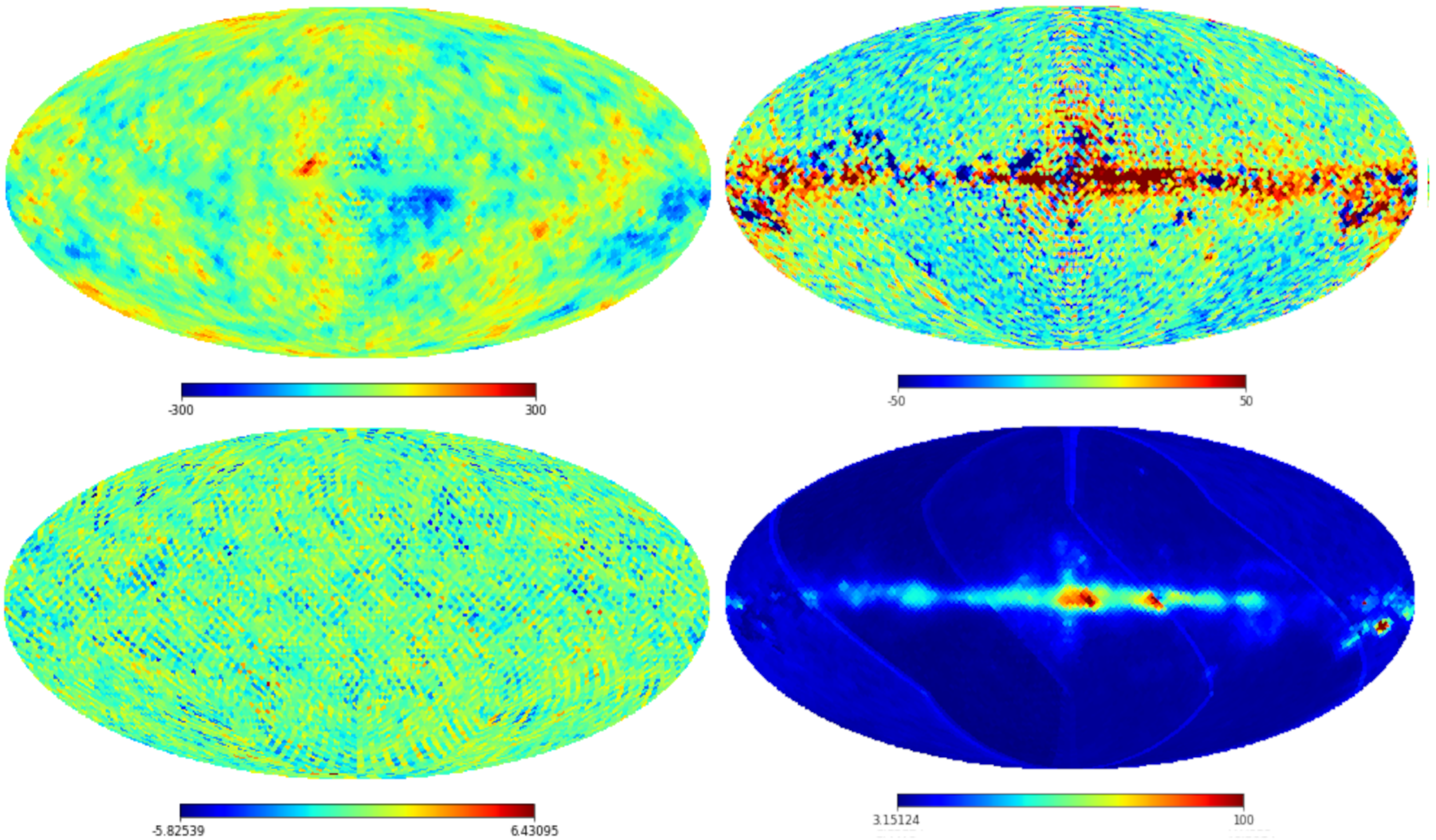}
\caption{In the top left panel we show the ${\tt mean\_CMap}$ obtained by our method 
after minimized the foregrounds present in the observed Planck maps at HealpPix 
$N_{\tt side}$ = 32 and with a $1.83^\circ$ beam smoothing.  
The difference between our ${\tt mean\_CMap}$ and the Planck COMMANDER 
map is shown in 
the top right panel. In bottom left we show the difference between  
our ${\tt mean\_CMap}$ map and the ${\tt bf\_CMap}$ obtained from the 
posterior distributions of the pixels while in the bottom right 
panel we show the Error map. }
\label{truemaps}
\end{figure}

\begin{figure}
 \centering
\includegraphics[scale=0.75]{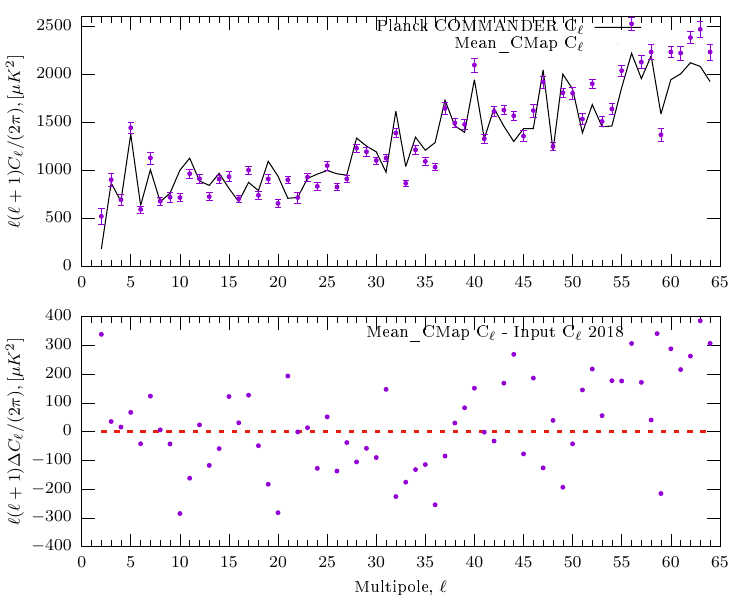}
\caption{The angular power spectrum estimated from the 
${\tt mean\_CMap}$ after performing wiener filetring is shown in the top panel    
with violet points.  We compare our power spectrum against the full-sky power 
spectrum estimated from the Planck COMMANDER map shown with black lines. 
Both the power spectra 
 agree well with each other in the lower multipoles but we see effect 
 of some residual contamination at higher multipoles. The difference between these 
 two power spectra is shown in the bottom panel.}
\label{datacl}
\end{figure}

\begin{figure}
 \centering
\includegraphics[scale=0.95]{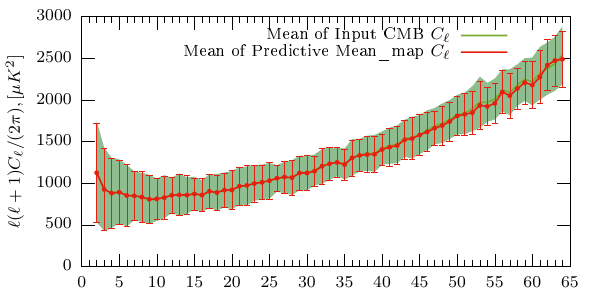}
\caption{We show the mean of all the 100 input CMB power spectra used in the testing 
dataset in green lines 
along with cosmic variance as light-green band. The mean of all the power spectra 
estimated from the Predictive ${\tt mean\_CMap}$ maps estimated after removing 
the foreground and noise contaminations in the testing dataset is shown in red line. 
The standard deviation error estimation by taking the standard deviation of all 
the 100 Predictive ${\tt mean\_CMap}$ maps. We see a very close match between 
both the power spectra and our estimated error is same as the cosmic variance induced 
error.}
\label{theocl}
\end{figure}

We utilize Planck 2018 released LFI 30, 44 and 70 GHz, 
HFI 100, 143, 217 and 353 GHz frequency maps to evaluate the 
performance of our trained ${\tt PUREPath}$ network. 
After properly taking care of the native beam and pixel effects in these maps, 
we preprocess the spherical full-sky maps as discussed in 
Section~\ref{preprocessing}. We follow the procedure outlined in Section~\ref{methodology} 
to estimate predictive CMB posterior, predictive mean (${\tt mean\_CMap}$), and 
Error map using Eqns.~\ref{pmean} and~\ref{pvar} respectively. 
In Figure~\ref{post_data_pix} we show 
the normalized probability
densities after dividing by the corresponding mode of
the marginalized density function for some selected pixels
over the sky. Using the mode values corresponding to each pixels probability 
densities we form a best-fit map (${\tt bf\_CMap}$). 
We show the ${\tt mean\_CMap}$ 
in the top left panel of  Figure~\ref{truemaps}. 
We compare our ${\tt mean\_CMap}$ with the publicly 
available Planck COMMANDER map which is downgraded to 
HealPix N$_{side}$ =  32 and smoothed  
by a Gaussian beam of FWHM = $1.83^\circ$ after properly accounting 
for the its native beam and pixel window functions. 
We display the  difference map obtained after 
taking the difference between our predicted mean cleaned CMB map and the 
Planck COMMANDER CMB map in the upper left panel.
The difference map reveals that the CMB map 
recovered by our ${\tt PUREPath}$  is in good agreement with the Planck COMMANDER 
CMB map at high Galactic latitudes. 
However, some discrepancies are observed near the low Galactic latitudes, 
especially close to the Galactic plane, due to high levels of 
thermal dust emissions. 
The difference between  ${\tt bf\_CMap}$  and  ${\tt mean\_CMap}$  
is shown in the bottom left panel of Figure~\ref{truemaps}.  
The lower right panel displays the Error map obtained. 
As expected the error map shows larger error near the galactic region 
while the error is quite low towards the higher latitudes. 

We estimate the full-sky power spectrum corresponding to  
our ${\tt mean\_CMap}$ map and is shown as violet points in the top panel of 
Figure~\ref{datacl}. We 
show the error bars corresponding 
our power ${\tt mean\_CMap}$ power spectrum estimated using the power spectra 
from different $\omega_i$s realizations. We see the error is minor and does not show 
any signs of noise at higher multipoles $\ell$. 
The full-sky power spectrum is obtained after performing wiener 
filtering using the full-sky power spectra estimated from the testing samples. 
The Planck COMMANDER full-sky spectrum is shown in black line and we note that our  
at power spectra matches well with the Planck COMMANDER spectrum. 
We show the difference between out power spectra and the Planck spectrum in the bottom 
panel. 

IN Figure~\ref{theocl} we show the mean of all the 100 power spectra estimated 
from the ${\tt mean\_CMap}$ corresponding to the 100 sets of testing data in 
red line. The mean of all the input CMB power spectra used in the testing data is 
shown in green line. The green band corresponds to the cosmic variance. 
We see a very close match between both the power spectrum. 
We also show the standard deviation error estimated from our predicted mean map 
power spectra. The very close match between the two errors signifies 
under the circumstance where we have a very good understanding of the nature of 
foregrounds and instrumental characteristics
our trained network is capable of minimizing foregrounds and noise effectively in 
the CMB observations. 
We show the detailed results from our testing phase, the CMB map and power spectra 
results from testing data in Appendix~\ref{testing}. 

\section{Salient Features of our Network}
\label{salient}
The network design of ${\tt PUREPath}$ facilitates in estimating  
a predictive CMB posterior conditioned on the input foreground contaminated 
CMB maps and the training dataset. Using this posterior,  
summary statistics like predictive mean, variance and other higher 
order statistics can be computed. 
The introduction of prior distributions over model 
parameters in our network acts as regularization. This is advantageous 
especially when the training data is 
limited or noisy. Bayesian networks are typically challenging to 
train due to computationally expensive complex calculations involved, and 
challenges encountered while scaling to large
datasets. However, use of Convolution2DFlipout layers, VI 
techniques  allows for efficient ways to approximate complex posterior distributions.
Currently our model minimizes foregrounds in the observed CMB maps at  
HealPix  $N_{side} = 32$ but can be 
scaled to higher resolutions.  
During inference, the Error map quantifies our model's uncertainty  
on the predicted cleaned CMB 
map. This can be particularly useful in contexts like using our 
cleaned map for cosmological 
parameter estimation and other cosmological analyses where 
understanding the uncertainty in the predicted map is as crucial as the 
predictions themselves.

In the current work, we divide the 
sky into 4 regions following specific scheme, but this can be modified to 
accommodate data from other full-/cut-sky observations. The composition 
of multi-encoder-latent-decoder set up for each region can be tailored 
according to requirement i.e., add more encoders if there are more 
input maps for some specific region or add more layers/filters etc 
for in-depth learning. 

\section{Conclusions \& Discussions}
\label{Conclusion}
By incorporating Bayesian machine learning techniques in conjunction 
with a U-Net and ResNet architecture, our model ${\tt PUREPath}$, 
offers a powerful framework for estimating the CMB posterior conditioned 
on the input data and the training dataset. Using this predictive CMB 
posterior we can estimate all the summary statistics like predictive 
mean cleaned CMB map, its per-pixel error estimate a standard deviation map etc. 
We leverage the Bayesian models inherent capability to quantify the uncertainty 
in its predictions, the per-pixel uncertainty estimates provided by our approach 
is crucial for understanding the confidence in predictions and 
for subsequent cosmological analyses. 

We train our model by simulating 1200 sets of foreground contaminated CMB maps 
at first 7 Planck frequency channels. We use thermal dust, synchrotron, free-free 
and AME as the major sources of foreground contaminations. 
Once trained, we implement our network using the observed 
CMB data provided by Planck satellite mission, we see 
that the our estimated predictive 
mean map matches quite well with the  Planck COMMANDER CMB map with some
differences in the Galactic region. 

Furthermore, the Bayesian framework offers additional advantages like including regularization through KL-Divergence, 
robustness to noise in the data etc. Currently 
our network uses maps at HealPix $N_{\tt side} = 32$ but 
can be scaled higher to handle maps at higher pixel resolution. 
The current sky-divisions are based on choice but can be tailored to 
include maps from say ground-based missions. 
To incorporate additional maps for any specific region, more encoder units 
needs to be added to process those maps in the corresponding U-Net. 
Our network enables detailed learning 
for any region by either increasing the layers or filters, 
 or by adding more ResNet units in the U-Net corresponding to that 
region while keeping other regions U-Net configuration same. 

Overall, the proposed approach provides a comprehensive solution for 
predictive CMB posterior estimation, enabling more accurate measurements 
of cosmological 
parameters and other cosmological analyses. 

\section{Acknowledgments}
 
This  work is based on observations obtained with Planck (http://www.esa.int/Planck). 
Planck is 
an ESA science mission with instruments and contributions directly funded by 
ESA Member States, 
NASA, and Canada.   
We acknowledge the use of Planck Legacy Archive (PLA) and the Legacy Archive 
for Microwave Background 
Data Analysis (LAMBDA). We  use publicly available HEALPix~\cite{Gorski:2004by} package  
(http://healpix.sourceforge.net) for the analysis of this work.  The network 
we have developed is based on 
the libraries provided by Tensorflow.

\bibliography{ms.bib}{}
\bibliographystyle{aasjournal}

\appendix

\section{Testing Results}
\label{testing}
From each set of model parameters $\omega_i$ sampled from the 
posterior $P(\omega|\mathcal{D})$ at every forward pass during inference,  
we estimate a predictive mean cleaned CMB map (either by sampling cleaned CMB 
maps using ${\tt tf.model(inputs).sample(n_{\tt sample})}$ and averaging them or 
by allowing the model to provide the distribution corresponding to the cleaned map through 
${\tt output = tf.model(inputs, training=False)}$ and estimating the mean and 
standard devaiation as follows:
\begin{equation}
{\tt mean = 
output.mean()}\, , \\ 
{\tt variance = 
output.variance()}\, , \\ 
{\tt and\,\, std = tf.sqrt(variance)}\, .    
\end{equation} 
We repeat this  10000 times.  
\subsection{CMB Posterior}
\begin{figure}
 \centering
\includegraphics[scale=0.185]{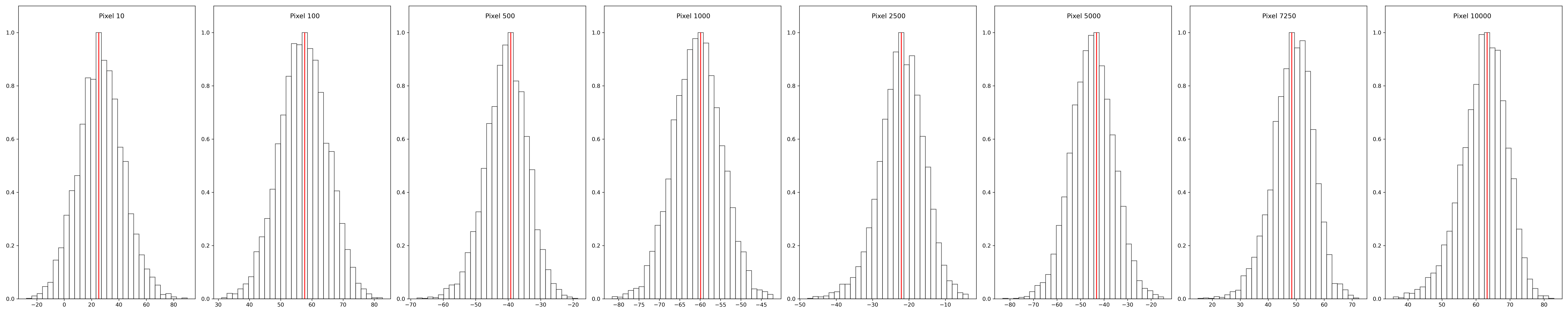}
\caption{We show the normalized densities of the CMB pixel temperatures for some 
selected pixels in black. The normalization for each density is
such that the peak corresponds to a value of unity. 
The position of mode value is shown by the red vertical lines.}
\label{post_sim_pix}
\end{figure}

The sampled predictive mean cleaned CMB maps corresponding to each set 
of $\omega_i$ is used 
to estimate the CMB posterior distribution ${\hat P}({\bf y}^\ast|{\bf x}^\ast, \mathcal{D})$ 
following Eqn.~\ref{posterior} for a given set 
of input ${\bf x}^\ast$.  
In Figure~\ref{post_sim_pix} we show 
the normalized probability
densities after dividing by the corresponding mode of
the marginalized density function for some selected pixels
over the sky. These density functions are
approximately symmetric. This estimated predictive CMB posterior corresponds to 
one set of foreground contaminated maps. Similarly, we can estimate the posterior 
density corresponding to every other set of simulated foregrounds contaminated 
maps in the testing dataset.  
\subsection{CMB Maps \& Angular Power Spectrum}

\begin{figure}
\centering
\includegraphics[scale=0.3]{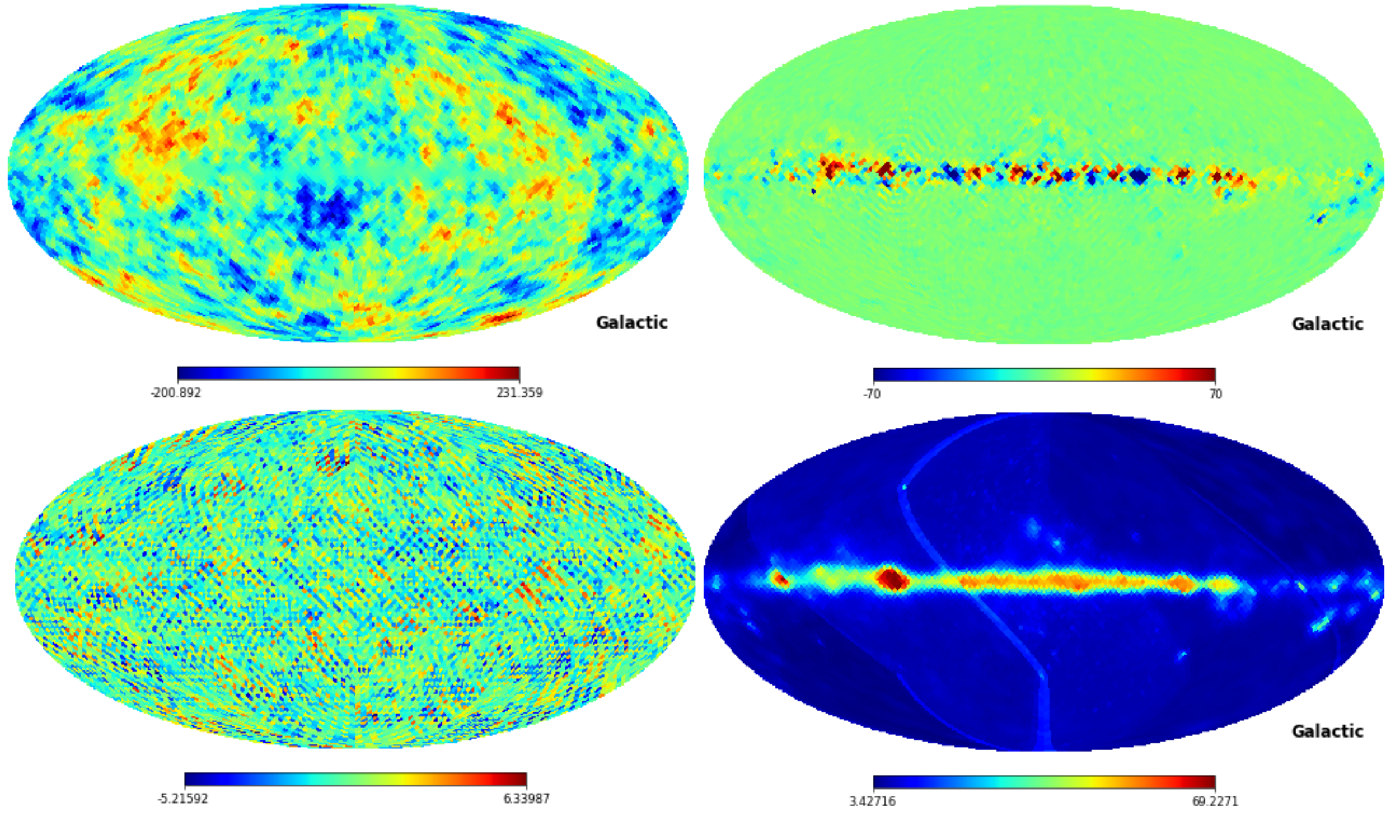}
\caption{In the top left panel we show the ${\tt mean\_CMap}$ obtained by our method.
The difference between our ${\tt mean\_CMap}$ and the input CMB map used is shown in 
the top right panel. In bottom left and right panels we show the difference of 
${\tt mean\_CMap}$ map from ${\tt best\_CMap}$ and the Error map respectively. }
\label{predmaps}
\end{figure}

\begin{figure}
 \centering
\includegraphics[scale=0.8]{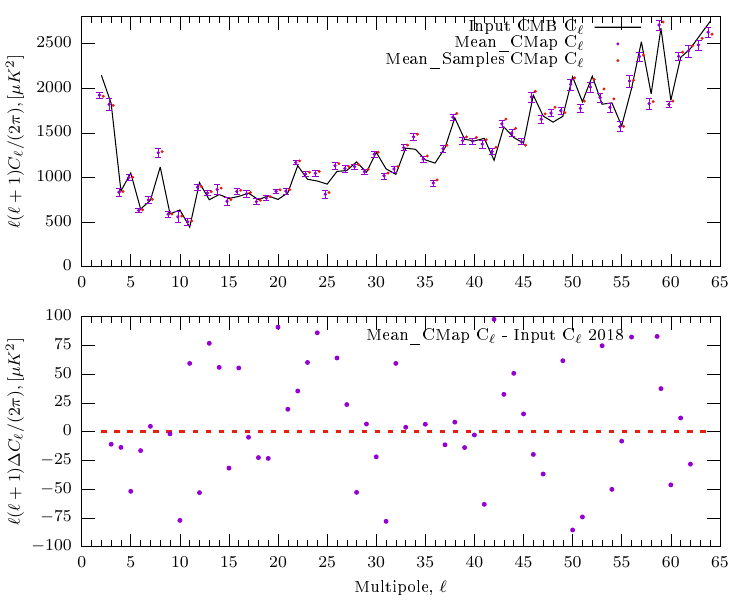}
\caption{We show the angular power spectrum estimated from the ${\tt mean\_CMap}$ 
with violet points while the input CMB power spectrum is shown in 
black line. We see both the power spectra agree with each other. The power spectrum 
obtained after taking the mean of 10000 power spectra corresponding to the 
sampled mean maps is shown in red points. }
\label{predcl}
\end{figure}
 We show our ${\tt mean\_CMap}$ in the top left panel of 
Figure~\ref{predmaps}. The difference between our ${\tt mean\_CMap}$  
and the input CMB used is displayed in the top right panel of Figure~\ref{predmaps}. The 
residual map shows that our ${\tt mean\_CMap}$ matches very well with the
CMB realization used in the simulation and our model is capable of recovering  
CMB signal quite well. 
The difference between the ${\tt bf\_CMap}$ and  the ${\tt mean\_CMap}$ 
is shown in the 
bottom left panel of Figure~\ref{predmaps} and the 
per-pixel error estimate corresponding to the ${\tt mean\_CMap}$  
in the bottom right panel. 
The error map represents the standard 
deviation of the predictive posterior distribution of the CMB map. 

Using the  ${\tt mean\_CMap}$ we estimate 
the angular power spectrum after 
properly accounting for the beam and pixel effects and is shown in Figure~\ref{predcl} 
using violet points. The input CMB power spectrum is shown in black line. We 
see that our  ${\tt mean\_CMap}$ power spectrum matches well with the 
input CMB power spectrum.  
From both the map and power spectrum comparisons we see that our  
model is able to accurately reconstruct a cleaned CMB map 
after minimizing the foregrounds 
present in the input simulated maps.

\end{document}